\documentclass[aps,prl,reprint,longbibliography,superscriptaddress]{revtex4-2}
\usepackage{mathrsfs}
\usepackage{amsmath,gensymb}
\usepackage{amsfonts}
\usepackage{amssymb}
\usepackage{amsthm}
\usepackage{graphicx}
\usepackage{natbib}
\usepackage{color}
\usepackage[
pdffitwindow=true,
colorlinks=true,
frenchlinks=false,
linkcolor=blue,
anchorcolor=blue,
citecolor=blue,
filecolor=blue,
urlcolor=blue,
bookmarks=true,
bookmarksopen=true,
bookmarksnumbered=true,
bookmarksopenlevel=1,
plainpages=false,
pdfpagelabels=true,
breaklinks
]{hyperref}
\usepackage{bm}
\usepackage[caption=false]{subfig}
\usepackage{verbatim}
\usepackage[per-mode=symbol,separate-uncertainty]{siunitx}

\begin{document}
	\title{Observation of nonreciprocal magnon Hanle effect}
	
	\author{Janine~G{\"u}ckelhorn}
	\email{janine.gueckelhorn@wmi.badw.de}
	\affiliation{Walther-Mei{\ss}ner-Institut, Bayerische Akademie der Wissenschaften, D-85748 Garching, Germany}
	\affiliation{Physik-Department, Technische Universit\"{a}t M\"{u}nchen, D-85748 Garching, Germany}
	
	\author{Sebasti{\'a}n de-la-Pe\~{n}a}
	\affiliation{Condensed Matter Physics Center (IFIMAC) and Departamento de F\'{i}sica Te\'{o}rica de la Materia Condensada, Universidad Aut\'{o}noma de Madrid, E-28049 Madrid, Spain} 
	
	\author{Matthias~Grammer}
	\affiliation{Walther-Mei{\ss}ner-Institut, Bayerische Akademie der Wissenschaften, D-85748 Garching, Germany}
	\affiliation{Physik-Department, Technische Universit\"{a}t M\"{u}nchen, D-85748 Garching, Germany}
	
	\author{Monika~Scheufele}
	\affiliation{Walther-Mei{\ss}ner-Institut, Bayerische Akademie der Wissenschaften, D-85748 Garching, Germany}
	\affiliation{Physik-Department, Technische Universit\"{a}t M\"{u}nchen, D-85748 Garching, Germany}
	
	\author{Matthias~Opel}
	\affiliation{Walther-Mei{\ss}ner-Institut, Bayerische Akademie der Wissenschaften, D-85748 Garching, Germany}
	
	\author{Stephan~Gepr{\"a}gs}
	\affiliation{Walther-Mei{\ss}ner-Institut, Bayerische Akademie der Wissenschaften, D-85748 Garching, Germany}
	
	\author{Juan Carlos Cuevas}
	\affiliation{Condensed Matter Physics Center (IFIMAC) and Departamento de F\'{i}sica Te\'{o}rica de la Materia Condensada, Universidad Aut\'{o}noma de Madrid, E-28049 Madrid, Spain} 
	
	\author{Rudolf~Gross}
	\affiliation{Walther-Mei{\ss}ner-Institut, Bayerische Akademie der Wissenschaften, D-85748 Garching, Germany}
	\affiliation{Physik-Department, Technische Universit\"{a}t M\"{u}nchen, D-85748 Garching, Germany}
	\affiliation{Munich Center for Quantum Science and Technology (MCQST), D-80799 M\"{u}nchen, Germany}
	
	\author{Hans~Huebl}
	\affiliation{Walther-Mei{\ss}ner-Institut, Bayerische Akademie der Wissenschaften, D-85748 Garching, Germany}
	\affiliation{Physik-Department, Technische Universit\"{a}t M\"{u}nchen, D-85748 Garching, Germany}
	\affiliation{Munich Center for Quantum Science and Technology (MCQST), D-80799 M\"{u}nchen, Germany}
	
	\author{Akashdeep Kamra}
	\email{akashdeep.kamra@uam.es}
	\affiliation{Condensed Matter Physics Center (IFIMAC) and Departamento de F\'{i}sica Te\'{o}rica de la Materia Condensada, Universidad Aut\'{o}noma de Madrid, E-28049 Madrid, Spain}
	
	\author{Matthias~Althammer}
	\email{matthias.althammer@wmi.badw.de}
	\affiliation{Walther-Mei{\ss}ner-Institut, Bayerische Akademie der Wissenschaften, D-85748 Garching, Germany}
	\affiliation{Physik-Department, Technische Universit\"{a}t M\"{u}nchen, D-85748 Garching, Germany}

	\begin{abstract}
		The precession of magnon pseudospin about the equilibrium pseudofield, the latter capturing the nature of magnonic eigen-excitations in an antiferromagnet, gives rise to the magnon Hanle effect. Its realization via electrically injected and detected spin transport in an antiferromagnetic insulator demonstrates its high potential for devices and as a convenient probe for magnon eigenmodes and the underlying spin interactions in the antiferromagnet. Here, we observe a nonreciprocity in the Hanle signal measured in hematite using two spatially separated platinum electrodes as spin injector/detector. Interchanging their roles was found to alter the detected magnon spin signal. The recorded difference depends on the applied magnetic field and reverses sign when the signal passes its nominal maximum at the so-called compensation field. We explain these observations in terms of a spin transport direction-dependent pseudofield. The latter leads to a nonreciprocity, which is found to be controllable via the applied magnetic field. The observed nonreciprocal response in the readily available hematite films opens interesting opportunities for realizing exotic physics predicted so far only for antiferromagnets with special crystal structures.
	\end{abstract}

	\maketitle
	
	
	The quantized excitations of the spin system in ordered magnets - magnons - offer a unique platform for intriguing science and technology. Their solid state host and associated quantized spin make them promising as information carriers~\cite{Kajiwara2010,Kruglyak2010,Bauer2012,Chumak2015}, comparable to electrons. At the same time, their bosonic nature allows for phenomena typically exploited in, e.g., optics and optomechanics~\cite{Aspelmeyer2014}. Combining these features, antiferromagnetic magnons with their high frequencies~\cite{Keffer1952,Jungwirth2016,Baltz2018} and tunable spin~\cite{Kamra2017,Liensberger2019} offer fast operation and robustness against thermal fluctuations, among several advantages~\cite{Jungwirth2016,Baltz2018}. Numerous theoretical proposals exploit the diversity and engineerability of antiferromagnetic magnons for unprecedented phenomena~\cite{Rezende2019,Lan2017,Mook2014,Cheng2016,Zyuzin2016,Cheng2016B,Daniels2018,Kawano2019,Kawano2019B}. However, their high frequencies also pose challenges making conventional GHz spectroscopies,
	although showing rapid experimental progress with ferromagnetic magnons, not suitable for probing antiferromagnetic magnons.
	
	Electrically injected and detected magnonic spin transport overcomes this limit to a large extent and is mediated by magnons in the full frequency range~\cite{Cornelissen2015,Goennenwein2015,Zhang2012,Li2016,Velez2016,Cornelissen2016,Lebrun2018,Han2020,Lebrun2020,Schlitz2021}. In this respect, the recent observation of the magnon Hanle effect in an antiferromagnetic insulator has opened new opportunities~\cite{Wimmer2020,Kamra2020,Gueckelhorn2022,Ross2020}. On the one hand, it shows that the antiferromagnetic magnon pseudospin can be manipulated and used in devices, similar to the electronic spin in spintronics~\cite{Kikkawa1999,Jedema2002}. On the other hand, it offers a new powerful tool for studying the rich nature of antiferromagnetic magnons, parameterized via a pseudospin, thereby providing crucial information about the underlying spin interactions~\cite{Kawano2019,Kawano2019B,Cheng2016,Shen2020,Shen2021}.
	
	Several of the exciting theoretical predictions in quantum matter, such as topological antiferromagnetic magnons~\cite{Mook2014,Cheng2016B,Zyuzin2016,Owerre2016}, essentially exploit inversion symmetry-breaking in the spin system. This, in turn, is intricately related to spin-orbit interaction and nonreciprocity~\cite{Tokura2018}. These have been the basis of exciting chiral or rectification phenomena observed across platforms~\cite{Tokura2018}, from supercurrents~\cite{Ando2020} to magnetoacoustic waves~\cite{Kuess2020,Kuess2022}. Moreover, nonreciprocal antiferromagnetic magnons have been observed in $\alpha$-$\mathrm{Cu}_2\mathrm{V}_2\mathrm{O}_7$ using neutron scattering experiments~\cite{Gitgeatpong2017,Santos2020}. However, antiferromagnetic materials with broken inversion symmetry are still scarce and mostly do not offer high N\'{e}el temperatures. This is in contrast to ferromagnets, where nonreciprocal magnonic responses have been reported in various widely used magnetic hybrids employing easily accessible detection schemes at room temperature, thereby triggering rapid advancements~\cite{Tokura2018,Moon2013,Kim2016,Nembach2015,Garst2017,Wang2020,Yu2021,Schlitz2021}. Hence, several undiscovered nonreciprocity-based phenomena with antiferromagnetic magnons await finding suitable widely available materials and detection methods.

	\begin{figure}[t]
		\centering
		\includegraphics[width=85mm]{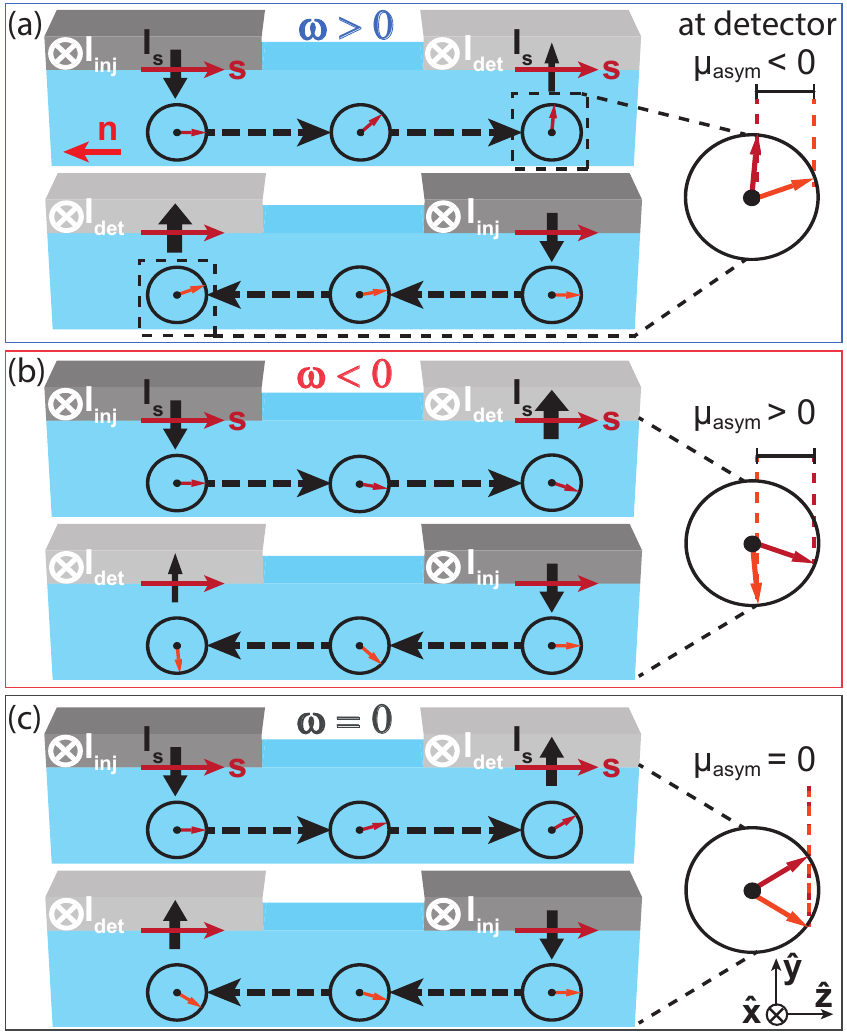}
		\caption{Schematic depiction of magnon spin and pseudospin transport in an antiferromagnetic insulator (AFI)~\cite{Wimmer2020,Kamra2020}, shown in blue. Two normal metal (NM) leads act as injector (dark gray) and detector (light gray) of magnonic spin current, which corresponds to the pseudospin $z$-component in the AFI. Due to easy-plane anisotropy in the studied AFI, the pseudofield is directed along $\hat{\pmb{x}}$ and the pseudospin precesses about this direction as the magnons propagate from injector to detector. Due to slightly different pseudofields in the forward (upper panels) and backward (lower panels) propagation directions, there is a difference $\mu_{\mathrm{asym}}$ in the observed magnon signal ($\propto \mu_{sz}$) which is (a) positive, (b) negative, and (c) zero for the corresponding average pseudofield $\omega$. This sign change in the difference is qualitatively understood from the different pseudofield precession rates as captured in the depiction here.}
		\label{Fig1}
	\end{figure}

	Here, we report a nonreciprocal response in the widely available antiferromagnetic insulator (AFI) hematite, observed as direction-dependent electrically-induced magnon spin transport and Hanle effect~\cite{Wimmer2020}. The degree of nonreciprocity is found to vary with the equilibrium N\'{e}el vector direction and thus, is tunable. Our observations are understood in terms of the different pseudofields, and thus pseudospin precession rates, experienced by magnons propagating in the forward and backward directions. Since the pseudofield is directly related to the magnon eigenmodes and the underlying spin interactions~\cite{Kamra2020}, our observation provides clear evidence for the presence of inversion symmetry breaking in the AFI/substrate system. Furthermore, the observed inversion-asymmetric pseudofield demonstrates the existence of an emergent magnon pseudospin-orbit interaction~\cite{Kawano2019B,Shen2020,Santos2020}. 
	
	Before delving into the theoretical details, we briefly describe the sample configuration and qualitatively discuss the underlying physics~\cite{Wimmer2020,Kamra2020}. Our device consists of an AFI film (hematite) upon which two spatially separated normal metal (NM) electrodes (Pt) have been deposited (see Fig.~\ref{Fig1}). Spin-orbit interaction in the NMs causes spin-charge coupling and allows for electrical injection and detection of magnon spin currents~\cite{Hirsch1999,Saitoh2006,Sinova2015}. Let us consider the upper panel of Fig.~\ref{Fig1}(a). The left NM electrode injects a magnonic spin current into the AFI, which corresponds to injecting a $z$-polarized pseudospin current. The AFI under consideration bears an easy-plane anisotropy, which harbors $x$-directed pseudofield and, correspondingly, spin-0 magnon eigenmodes. Thus, the spin-1 magnons injected by the NM are not the eigenmodes and start to transmute into other kinds of magnons with varying spin. This process is represented by pseudospin precession about the pseudofield $\omega_+\hat{\pmb{x}}$. Consequently, the magnon spin, given by the pseudospin $z$-component, detected by the right NM depends on the pseudofield $\omega_+$. The latter can further be controlled via an applied magnetic field~\cite{Wimmer2020} and vanishes at a specific value denoted as $H_{\mathrm{c}}$. 
	
	If we interchange the two NM electrodes roles' [lower panel, Fig.~\ref{Fig1}(a)] injecting spin with the right and detecting it using the left, magnons may experience a slightly different pseudofield $\omega_{-}\hat{\pmb{x}}$ due to inversion symmetry breaking. Consequently, the magnon spin signal detected in this configuration is slightly different. This difference ($\propto \delta \omega \equiv (\omega_+ - \omega_-)/2$) allows us to quantify the pseudofield nonreciprocity in the system. Furthermore, as depicted in Fig.~\ref{Fig1}(b), this difference changes sign together with the average pseudofield $\omega \equiv (\omega_+ + \omega_-)/2$ due to a corresponding reversal of the precession sense. Moreover, this difference vanishes with the pseudofield, as depicted schematically in Fig.~\ref{Fig1}(c). These key features, validated by our experimental data reported below, allow us to confirm the nonreciprocal pseudofield as the origin of the observed nonreciprocity in the magnon Hanle effect.

	The nonequilibrium magnons and their transport in the AFI can be described in terms of the pseudospin chemical potential $\pmb{\mu}_s$, which is a vector~\cite{Kamra2020}. Its magnitude, $z$-component, and direction respectively capture the densities of nonequilibrium magnons, spin, and their nature. For the system of interest, it suffices to consider that $\pmb{\mu}_s$ varies only along the $z$-coordinate between the injector and detector (Fig.~\ref{Fig1}). It is thus described by a one-dimensional diffusion equation~\cite{Kamra2020,Gueckelhorn2022}:
	\begin{align}\label{eq:diff}
		\frac{\partial \pmb{\mu}_s}{\partial t} & = D_m \frac{\partial^2\pmb{\mu}_s}{\partial z^2} - \frac{\pmb{\mu}_s}{\tau_m} + \pmb{\mu}_s \times \omega \hat{\pmb{x}} - l \frac{\partial \pmb{\mu_s}}{\partial z} \times \delta \omega \hat{\pmb{x}},
	\end{align} 
	where $D_m$ is the diffusion coefficient, $\tau_m$ is the spin relaxation time, and $l$ is the mean free path, all quantities pertaining to the AFI magnons. The last term on the right hand side of Eq.~\eqref{eq:diff} is the new contribution here as compared to the previous inversion-symmetric considerations~\cite{Kamra2020,Gueckelhorn2022}. It is obtained by allowing different pseudofields in the forward ($+$) and backward ($-$) directions within the random walk model describing the diffusive pseudospin transport with precession~\cite{Kamra2020,delaPena2022,Fabian2007}. 
	
	The magnon spin injection by the NM is taken into account via the boundary conditions at the injector location~\cite{Kamra2020,Gueckelhorn2022} assumed to be $z = 0$: $- D_m \chi \partial \mu_{sz} / \partial z = + (-) j_{s0}$ and $\partial \mu_{sx,sy} / \partial z = 0$. Here, $\chi$ is the susceptibility that relates the pseudospin density with its chemical potential, and $j_{s0}$ is the magnitude of the magnon spin current density driven by the injector NM. Its direction is positive (negative) for transport along $\hat{\pmb{z}}$ ($- \hat{\pmb{z}}$), which further leads to the $+$ ($-$) sign in the boundary condition above. Finally, imposing the stability requirement $\pmb{\mu}_s(z \to \infty) = 0$ [$\pmb{\mu}_s(z \to -\infty) = 0$] for the forward [backward] case, we obtain the desired solutions to Eq.~\eqref{eq:diff} for $\pmb{\mu}_{s}(z)$ for both cases: forward $z>0$ and backward $z<0$. 
	
\begin{figure}[t]
		\centering
		\includegraphics[width=85mm]{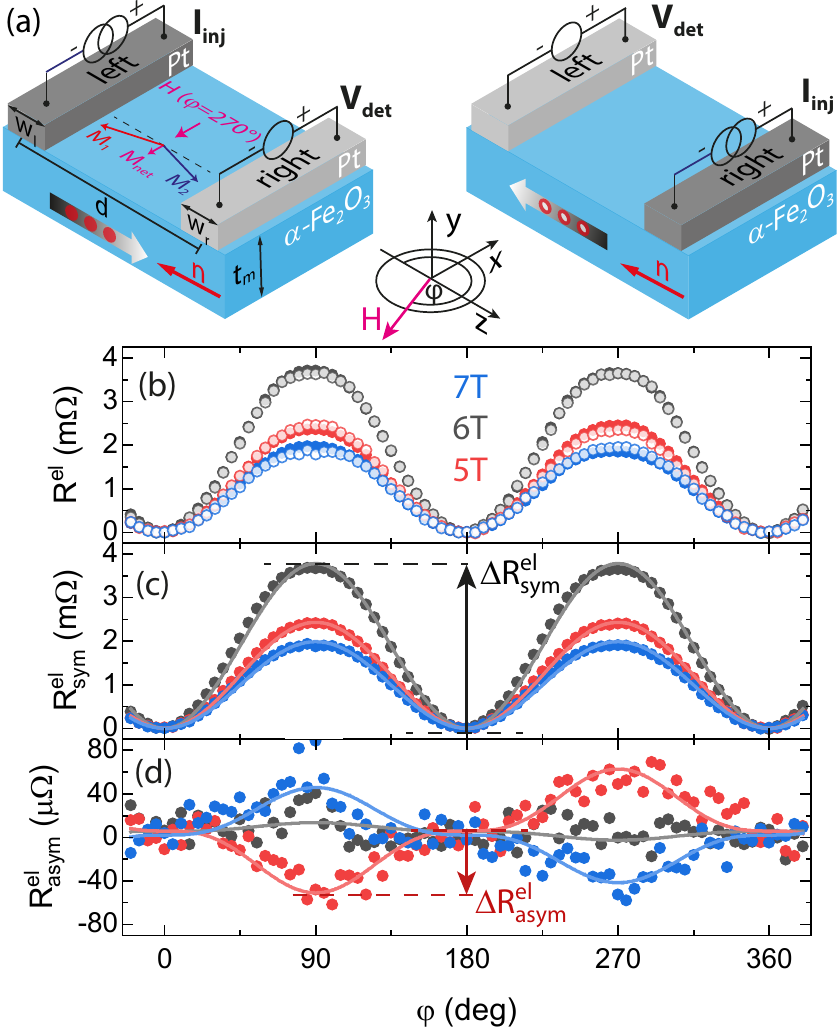}
		\caption{(a) Sketch of the sample configuration for the forward (left panel) and backward (right panel) transport directions, the electrical wiring scheme, and the coordinate system with the in-plane rotation angle $\varphi$ of the applied magnetic field $\mu_0\pmb{H}$. The corresponding net magnetization $\pmb{M}_\mathrm{net}$ is aligned along the applied magnetic field $\mu_0\pmb{H}$ ($\pmb{M}_\mathrm{net}\parallel \pmb{H}$), while the N\'{e}el order parameter $\pmb{n}\perp\pmb{H}$. (b) Angle dependence of the electrically induced magnon spin signal $R^\mathrm{el}\propto \mu_\mathrm{sz}$ measured at $T = \SI{250}{\kelvin}$ for a center-to-center distance of $d=\SI{1.2}{\micro\meter}$ and different magnetic field magnitudes for the $\SI{89}{\nano\meter}$ thick hematite film. The full and open circles, respectively, depict the measured signal in the forward and backward transport configurations. A constant offset arising from the experimental setup has been subtracted from the curves. (c) Symmetric part of the two measurement configurations for the same magnetic fields as in panel (b). The lines are fits to the expected $\Delta R_\mathrm{sym}^\mathrm{el}\sin^2(\varphi)$ function~\cite{Wimmer2020,Gueckelhorn2022}. (d) Antisymmetric part of the respective curves in panel (b). The lines represent a $\Delta R_\mathrm{asym}^\mathrm{el} \sin^3(\varphi)$ fit, indicating a $\sin \varphi$ dependence of the pseudofield nonreciprocity $\delta \omega$ as discussed in the text.}
		\label{Fig2}
	\end{figure}

	The detected magnon spin signal is directly proportional to $\mu_{sz}$ at the detector location~\cite{Kamra2020,Gueckelhorn2022}. Hence, $\mu_{sz}(+d)$ [$\mu_{sz}(-d)$] represents the detected magnon spin signal in the forward [backward] transport configuration (see upper [lower] panels in Fig.~\ref{Fig1}), where $d$ is the injector-detector distance [Fig.~\ref{Fig2}(a)]. We further define $\mu_{\mathrm{sym}} \equiv [\mu_{sz}(+d) + \mu_{sz}(-d)]/2$ and $\mu_{\mathrm{asym}} \equiv [\mu_{sz}(+d) - \mu_{sz}(-d)]/2$, which are evaluated as
	\begin{align}
		\mu_{\mathrm{sym}} & = \frac{l_\mathrm{m} j_{\mathrm{s}0} e^{- \frac{a d}{l_\mathrm{m}}}}{D_\mathrm{m} \chi \left(a^2 + b^2\right)}  \left[a \cos \left( \frac{b d}{l_\mathrm{m}} \right) -b \sin \left( \frac{b d}{l_\mathrm{m}} \right) \right], \nonumber \\
		\mu_{\mathrm{asym}} & = \frac{\omega  \delta \omega  \tau_m l}{|\omega| 2 l_m  } \frac{\partial \mu_{\mathrm{sym}}}{\partial b}, \label{eq:mu}
	\end{align}
	where $a \equiv  \sqrt{\left(1 + \sqrt{1 + \omega^2 \tau_m^2}\right)/2}$, $b \equiv  \sqrt{\left(-1 + \sqrt{1 + \omega^2 \tau_m^2}\right)/2}$, and $l_m \equiv \sqrt{D_m \tau_m}$ is the magnon diffusion length. Here, we have retained terms up to the first order in $l \delta \omega$, assuming $|l \delta \omega/ (l_m \omega)| \ll 1$.
	
	Equation \eqref{eq:mu} constitutes the desired and our main theoretical result, which is employed to analyze the experimental data below. The expression thus obtained for $\mu_{\mathrm{sym}}$ is the same as that in the previous inversion-symmetric analysis~\cite{Kamra2020}. In contrast, $\mu_{\mathrm{asym}}$ captures the nonreciprocity and is finite only when the pseudofield is nonreciprocal, i.e., when $\delta \omega \neq 0$. Furthermore, Eq.~\eqref{eq:mu} shows that $\mu_{\mathrm{asym}}$ manifests the odd-in-$\omega$ behavior motivated and discussed schematically in Fig.~\ref{Fig1}. This feature allows us to distinguish a nonreciprocal pseudofield contribution from other potential sources of nonreciprocity.
	

	In our experiments, we use a $t_\mathrm{m} = \SI{89}{\nano\meter}$ thick film of hematite ($\alpha-\mathrm{Fe}_2\mathrm{O}_3$) as the AFI. The hematite film undergoes a transition from an easy-axis to an easy-plane AFI above the Morin transition $T_\mathrm{M}\approx\SI{200}{\kelvin}$~\cite{Morin1950} (see Supplemental Material (SM)~\footnote{\label{fn:SM} See Supplemental Material at [url], which contains information on the sample fabrication and magnetometry measurements, details on the measurement procedure and additional data on the reciprocal effect for an $\SI{19}{\nano\meter}$ thin hematite film as well as a rigorous description of the fitting routine. It contains Ref.\,\cite{Ross2020_2}.} for details). All measurements are conducted in the easy-plane phase at $T=\SI{250}{\kelvin}$, where the hematite film features an out-of-plane Dzyaloshinskii-Moriya interaction (DMI), in agreement with previous works~\cite{Wimmer2020,Gueckelhorn2022}. As depicted in the left panel of Fig.~\ref{Fig2}(a), the equilibrium N\'{e}el vector $\pmb{n}$, the two sublattice magnetizations $\pmb{M}_1$ and $\pmb{M}_2$ as well as the induced net magnetization $\pmb{M}_\mathrm{net}=\pmb{M}_1+\pmb{M}_2$ due to DMI lie in the $xz$- or (0001) $\mathrm{Fe}_2\mathrm{O}_3$-plane. 
	Both, $\pmb{n}$ and $\pmb{M}_\mathrm{net}$ can be controlled by the orientation and magnitude of the applied magnetic field $H$, where $\pmb{M}_\mathrm{net}$ encodes the canting angle~\cite{Wimmer2020}. To investigate the magnon spin transport by all-electrical means, we employ two spatially separated narrow Pt strips on top of the film (see SM~\ref{fn:SM}\footnotemark[1] for fabrication details). 
	
	To characterize the sample we perform angle-dependent electrical transport measurements by changing the orientation of the external magnetic field $H$ within the $xz$-plane [see Fig.~\ref{Fig2}(a)]. A dc charge current $I_\mathrm{inj}=\SI{500}{\micro\ampere}$ is applied first to the left electrode leading to spin injection into the hematite film via the spin Hall effect (SHE)~\cite{Hirsch1999,Saitoh2006,Sinova2015}. The resulting diffusive pseudospin magnon current is detected electrically as a voltage signal $V_\mathrm{det}$ at the right electrode [left panel, Fig.~\ref{Fig2}(a)] (see SM~\ref{fn:SM}\footnotemark[1] for experimental details). In a second step, we interchange the injector and detector electrode, i.e. $I_\mathrm{inj}$ is injected at the right electrode and the voltage $V_\mathrm{det}$ is detected at the left Pt strip [right panel, Fig.~\ref{Fig2}(a)]. The measured magnon spin signal $R^\mathrm{el}=V_\mathrm{det}/I_\mathrm{inj}$ is plotted in Fig.~\ref{Fig2}(b) versus the angle $\varphi$ of the applied in-plane magnetic field for three different magnitudes $\mu_0H$ for both configurations. The full circles correspond to the forward transport direction [$+d$, left panel of Fig.~\ref{Fig2}(a)], while open circles represent the backward direction [$-d$, right panel of Fig.~\ref{Fig2}(a)].
	Evidently, all curves appear to exhibit the $\sin^2(\varphi)$ angular dependence characteristic of a factor $\sin (\varphi)$ contributed by both of the injection and detection processes~\cite{Cornelissen2015,Goennenwein2015}. However, a careful examination shows that there are differences between the two propagation directions for $\mu_0H = \SI{5}{\tesla}$ and $\SI{7}{\tesla}$, predominantly at $\varphi=\SI{90}{\degree}, \SI{270}{\degree}$, where $R^\mathrm{el}$ is largest. This corresponds to $\pmb{n}\parallel \hat{\pmb{z}}$ or $\pmb{H}\parallel\hat{\pmb{x}}$ as $\pmb{H}\perp\pmb{n}$.
	
	To quantify this observation, we plot the symmetric $R_\mathrm{sym}^\mathrm{el}= [R^\mathrm{el}(+d)+R^\mathrm{el}(-d)]/2$ and antisymmetric $R_\mathrm{asym}^\mathrm{el}= [R^\mathrm{el}(+d)-R^\mathrm{el}(-d)]/2$ components of the magnon spin signal for the two measurement configurations in Fig.~\ref{Fig2}(c) and (d), respectively~\cite{Schlitz2021}. The angle dependence in Fig.~\ref{Fig2}(c) follows a simple $\Delta R_\mathrm{sym}^\mathrm{el}\sin^2(\varphi)$ behavior, where $\Delta R_\mathrm{sym}^\mathrm{el}$ is the amplitude of the symmetric magnon spin signal. As discussed below (cf. Fig.~\ref{Fig2}), this amplitude also exhibits the expected Hanle curve in agreement with previous reports~\cite{Wimmer2020,Gueckelhorn2022}, where an inversion-symmetric analysis was used~\cite{Kamra2020}. As expected, we observe $\Delta R_\mathrm{sym}^\mathrm{el} \propto \mu_\mathrm{sym}$ [Eq.~\eqref{eq:mu}].	Fig.~\ref{Fig2}(d) shows that $R_\mathrm{asym}^\mathrm{el}$ is vanishingly small at $\mu_0H=\SI{6}{\tesla}$ over the whole angle range, while we observe a clear angle-dependence for $\mu_0H=\SI{5}{\tesla}$ and $\SI{7}{\tesla}$. However, the nonreciprocal signal $R_\mathrm{asym}^\mathrm{el} \propto \mu_{\mathrm{asym}}$ [Eq.\eqref{eq:mu}] is about two orders of magnitude smaller. Moreover, it follows a $\Delta R_\mathrm{asym}^\mathrm{el}\sin^3(\varphi)$ dependence, with $\Delta R_\mathrm{asym}^\mathrm{el}$ denoting the amplitude [cf. Fig.~\ref{Fig2}(d)]. The $\sin^2(\varphi)$ dependence originates from the injection and detection process via the SHE, while the additional factor of $\sin(\varphi)$ is contributed by $\delta\omega$ as per Eq.~\eqref{eq:mu}. A similar nonreciprocity and angle-dependent $\delta \omega$ are also observed in other samples and structures with various distances $d$, for example in a $t_\mathrm{m} = \SI{19}{\nano\meter}$ thin hematite film (see SM~\ref{fn:SM}\footnotemark[1]). Fig.~\ref{Fig2}(d) also clearly shows the opposite signs of $R_\mathrm{asym}^\mathrm{el}$ for $\mu_0H=\SI{5}{\tesla}$ and $\SI{7}{\tesla}$. To analyze this behavior, we now examine the magnetic field dependence of the amplitudes. 
	
	\begin{figure}[t]
		\centering
		\includegraphics[width=85mm]{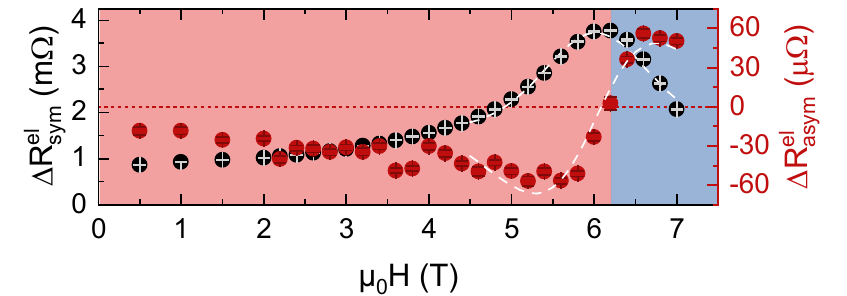}
		\caption{Magnetic field dependence of the symmetric and antisymmetric magnon spin signal $\Delta R_\mathrm{sym}^\mathrm{el}\propto\mu_{\mathrm{sym}}$ (black dots) and $\Delta R_\mathrm{asym}^\mathrm{el}\propto\mu_{\mathrm{asym}}$ (red dots), extracted from the data shown in Figs.~\ref{Fig2} (c) and (d), respectively. The dashed lines are fits to Eq.~\eqref{eq:mu}.
		}
		\label{Fig3}
	\end{figure}
	
	The quantities $\Delta R_\mathrm{sym}^\mathrm{el}$ and $\Delta R_\mathrm{asym}^\mathrm{el}$ are extracted from the fits in Figs.~\ref{Fig2}(c) and (d) and plotted in Fig.~\ref{Fig3} as a function of the magnetic field magnitude $\mu_0H$. Clearly, $\Delta R_\mathrm{sym}^\mathrm{el}$ (black dots) exhibits the expected Hanle curve [$\propto \mu_{\mathrm{sym}}$, Eq.~\eqref{eq:mu}] with a compensation field of $\mu_0H_\mathrm{c}=\SI{6.2}{\tesla}$~\cite{Wimmer2020,Gueckelhorn2022}. Moreover, $\Delta R_\mathrm{asym}^\mathrm{el}$ (red dots) approaches zero at $\mu_0H_\mathrm{c}$ and then changes sign for $\mu_0H>\mu_0H_\mathrm{c}$.	This behavior of the antisymmetric signal is consistent with the qualitative picture discussed above (Fig.~\ref{Fig1}) and confirms the antisymmetric pseudofield as the origin of the observed nonreciprocity. Furthermore, the magnetic field dependencies of $\Delta R_\mathrm{sym}^\mathrm{el}$ and $\Delta R_\mathrm{asym}^\mathrm{el}$ are reproduced well (white dashed lines in Fig.~\ref{Fig3}) via Eq.~\eqref{eq:mu} using a single consistent set of parameters. Here, we have restricted the fit range for $\Delta R_\mathrm{sym}^\mathrm{el}$ and $\Delta R_\mathrm{asym}^\mathrm{el}$ to $\mu_0H=4.5-\SI{7}{\tesla}$, as the simplified considerations resulting in Eq.~\eqref{eq:mu} do not account for low-energy magnons, which contribute to the spin transport at smaller magnetic fields~\cite{Gueckelhorn2022} (see SM~\ref{fn:SM}\footnotemark[1] for fit procedure and parameters).
	
	Thus, employing our theoretical analysis leading to Eq.~\eqref{eq:mu}, we deduce the origin of our experimentally observed nonreciprocity to be an antisymmetric pseudofield $\delta \omega$, finding it to be angle dependent as $\sin (\varphi)$. This angle dependence is reminiscent of a related, but distinct, nonreciprocity of the magnon dispersion found in yttrium iron garnet/gadolinium gallium garnet heterostructures~\cite{Schlitz2021}, which has been attributed to the interfacial DMI. Considering that the antisymmetric signal in our experiments is smaller in thicker samples (see SM), our observed nonreciprocity also likely stems from the interface between hematite and its substrate. This has not been observed before, to the best of our knowledge, and opens novel perspectives for engineering nonreciprocal effects in a widely employed AFI. At the same time, the spin Hamiltonians used to describe hematite in the literature~\cite{Wimmer2020}, for example in the original article by Moriya~\cite{Moriya1960}, could be oversimplified and might have missed such a nonreciprocity stemming from the crystal structure. To examine this potential origin, atomistic spin modeling of hematite taking into account its exact crystal structure is desirable and, hopefully, will be motivated by our findings~\cite{Ross2021,Brehm2022}.

	In summary, we demonstrate nonreciprocal magnon spin transport in the widely used antiferromagnetic insulator - hematite - employing electrical spin injection and detection. Our theoretical modeling allows us to understand this observation as due to an antisymmetric pseudofield along the spin transport direction. It further enables extraction of its dependence on the applied magnetic field. This antisymmetric pseudofield, in turn, directly translates to magnon dispersion~\cite{Kamra2020} and constitutes an observation of emergent pseudospin-orbit interaction. Hence, our work establishes nonlocal magnon transport as a powerful probe for underlying spin interactions in antiferromagnetic insulators. It also demonstrates the widely available hematite as a promising material for searching topological and nonreciprocal phenomena.

	\begin{acknowledgments}
		We gratefully acknowledge financial support from the Deutsche Forschungsgemeinschaft (DFG, German Research Foundation) under Germany's Excellence Strategy -- EXC-2111 -- 390814868 and project AL2110/2-1, and the Spanish Ministry for Science and Innovation -- AEI Grant CEX2018-000805-M (through the ``Maria de Maeztu'' Programme for Units of Excellence in R\&D).
	\end{acknowledgments}

\end{document}